\documentclass[prd,twocolumn,reprint,superscriptaddress,nofootinbib,preprintnumbers,longbibliography]{revtex4-1}

\usepackage[english]{babel}
\usepackage[utf8]{inputenc}
\usepackage{amssymb}
\usepackage{amsmath}
\usepackage{color}
\usepackage{hyperref}
\usepackage{bbm}
\usepackage{epsfig}
\usepackage{multirow}

\usepackage[normalem]{ulem}

\usepackage[dvipsnames]{xcolor}
\usepackage{xspace}

\newcommand{\e}{\varepsilon}
\newcommand{\s}{\sigma}

\newcommand{\up}{\uparrow}
\newcommand{\down}{\downarrow}
\newcommand{\w}{\omega}

\newcommand{\de}{{\rm d}}
\newcommand{\dd}{\partial}

\newcommand{\A}{\mathcal{A}}
\newcommand{\ket}[1]{\left| {#1} \right\rangle }

\newcommand{\GF}[1]{\langle\!\langle #1\rangle\!\rangle}
\newcommand{\vk}{\mathbf{k}}
\newcommand{\Jeff}{J^{\mathrm{eff}}}

\newcommand{\GS}[1]{\Gamma_{\mathrm{S}#1}}
\newcommand{\Gmin}{G_{\mathrm{min}}}
\newcommand{\Gmax}{G_{\mathrm{max}}}
\newcommand{\coef}{\mathcal{C}}

\renewcommand{\Im}{\mathrm{Im}}
\newcommand{\dk}{d^\dagger}
\newcommand{\ck}{c^\dagger}

\newcommand{\Sec}[1]{Sec.~\ref{sec:#1}}
\newcommand{\beq}{ \begin{equation} } 
\newcommand{\eeq}{ \end{equation} }
\newcommand{\beqa}{\begin{eqnarray}}
\newcommand{\eeqa}{\end{eqnarray}}
\newcommand{\nn}{\nonumber}
\newcommand{\es}{& = &}

\newcommand{\fig}[1]{Fig.~\ref{fig:#1}}
\newcommand{\figs}[1]{Figs.~\ref{fig:#1}}
\newcommand{\eq}[1]{Eq.~(\ref{#1})}
\newcommand{\eqs}[1]{Eqs.~(\ref{#1})}

\hypersetup{
    bookmarks=false,        
    colorlinks=true,        
    linkcolor=red,        
    citecolor=blue,        
    filecolor=blue,      
    urlcolor=blue
}

\begin{document}

\title{Nonlocal pairing as a source of spin exchange and Kondo screening}

\author{Krzysztof P. W{\'o}jcik}
\email{kpwojcik@ifmpan.poznan.pl}
\affiliation{Institute of Molecular Physics, Polish Academy of Sciences, 
			 Smoluchowskiego 17, 60-179 Pozna{\'n}, Poland}

\author{Ireneusz Weymann}
\affiliation{Faculty of Physics, Adam Mickiewicz University, 
			 Umultowska 85, 61-614 Pozna{\'n}, Poland}

\date{\today}

\begin{abstract}
We show that the Kondo screening in a correlated 
double quantum dot structure may be caused solely by the proximity of a superconductor,
which induces nonlocal pairing by Andreev reflection processes.
This leads to an effective exchange interaction,
which we estimate perturbatively and corroborate the analytical predictions
by the numerical renormalization group calculations,
using an effective model for the superconductor-proximized nanostructure.
We determine the dependence of the relevant Kondo temperature on the coupling to superconductor
and predict a characteristic modification of conventional low-temperature transport behavior,
which can be used to experimentally distinguish this phenomenon from other Kondo effects. 
The occurrence of nonlocal pairing exchange does not depend on details of the proposed setup,
therefore it can be also of relevance for the bulk materials, such as heavy-fermion compounds.
\end{abstract}

\maketitle

\section{Introduction}
\label{sec:Intro}

The exchange interactions control the magnetic order and properties of a vast number of materials
\cite{White2006Dec}
and lead to many fascinating phenomena, such as various types of the Kondo effect 
\cite{Kondo,NozieresBlandin,Pustilnik_Glazman}.
Double quantum dots (DQDs), and in general multi-impurity systems, constitute
a convenient and controllable playground,
where nearly as much different exchange mechanisms compete with each other to
shape the ground state of the system.
\emph{Local exchange} between the spin of a quantum dot (QD)
and the spin of conduction band electrons gives rise to the
Kondo effect \cite{Kondo,Hewson_book}. 
\emph{Direct exchange} arriving with an additional side-coupled QD may destroy it or lead to the 
two-stage Kondo screening \cite{Pustilnik_Glazman,Cornaglia,Granger,ZitkoBonca,ZitkoPRB2010,Ferreira}.
In a geometry where the two QDs contact the same lead, conduction band electrons 
mediate the \emph{RKKY exchange} \cite{RK,K,Y}. The RKKY interaction competes
with the Kondo effect and leads to the quantum phase transition of a still debated nature
\cite{Doniach,Jones,Affleck,Bork,Neel,KondoRKKYexp,Hans,Hans2,Fabian}.
Moreover, in DQDs coupled in series also \emph{superexchange} can alter the Kondo physics significantly
\cite{Zitko_2QDEx,Sela}.

Recently, hybrid quantum devices, in which the interplay between various magnetic correlations
with superconductivity (SC) plays an important role, have become an important direction of research
\cite{hybridQDs,SCspintronics}. In particular, chains of magnetic atoms on SC surface have proven 
to contain self-organized Majorana quasi-particles and exotic spin textures
\cite{Braunecker,Klinovaja,Vazifeh,Yazdani},
while hybrid DQD structures have been used to split the Cooper pairs coherently into two entangled 
electrons propagating to separated normal leads \cite{CPS1,CPS2,CPS4,CPS5,CPS9}.
The latter is possible due to non-local (\emph{crossed}) Andreev reflections (CARs),
in which each electron of a Cooper pair tunnels into different QD, and
subsequently to attached lead. Such processes give rise to an exchange mechanism \cite{Yao},
that we henceforth refer to as \emph{the CAR exchange}, which can greatly modify
the low-temperature transport behavior of correlated hybrid nanostructures.

The CAR exchange may be seen as RKKY-like interaction between
two nearby impurities on SC surface \cite{Yao}.
The effect can be understood as a consequence
of spin-dependent hybridization of the Yu-Shiba-Rusinov (YSR)
states \cite{Yu,Shiba,Rusinov} in SC contact,
caused both by the overlap of their wave functions
and their coupling to Cooper-pair condensate.
This process is the most effective when the YSR states 
are close to the middle of the SC gap, {\it e.g.} in the YSR-screened phase \cite{YSRscreening}.
The mechanism presented here is essentially the same,
yet in the considered regime can be understood
perturbatively without referring to YSR states,
as a consequence of the non-local pairing induced by SC electrode. 
In particular, the presence of YSR bound states close to the Fermi level 
is not necessary for significant consequences for the Kondo physics, 
as long as some inter-dot pairing is present.

The proximity of SC induces pairing in QDs \cite{RozhkovArovas,Buitelaar} 
and tends to suppress the Kondo effect if the superconducting energy gap $2\Delta$ 
becomes larger than the relevant Kondo temperature $T_K$ 
\cite{Buitelaar2002Dec,adatomsSC,Kondo_vs_SC1,Kondo_vs_SC2,Zitko_Kondo-Andreev,Zitko_S-QD-N,IW_Sau,YSRscreening}.
Moreover, the strength of SC pairing can greatly affect the Kondo physics in the sub-gap transport regime:
For QDs attached to SC and normal contacts, it can enhance the Kondo effect
\cite{DomanskiIW,KWIW,part1}, while
for DQD-based Cooper pair splitters, it tends to suppress both the $\mathrm{SU}(2)$ and $\mathrm{SU}(4)$ Kondo effects \cite{IW_Kacper}.
Our main result is that the non-local pairing induced by superconducting 
proximity effect, which gives rise to CAR exchange, can be the sole cause of the Kondo screening.
Moreover, relatively small values of coupling to SC, $\GS{}\ll U$, are sufficient for the effect to occur.
This is in contrast to the DQD system considered in Ref.~\cite{part1},
where only one of the quantum dots is proximized, such that 
CAR exchange cannot arise,
and the Kondo physics becomes qualitatively
affected only for $\GS{}\sim U/2$.%

\begin{figure}[bt]
\centering
\includegraphics[width=1\linewidth]{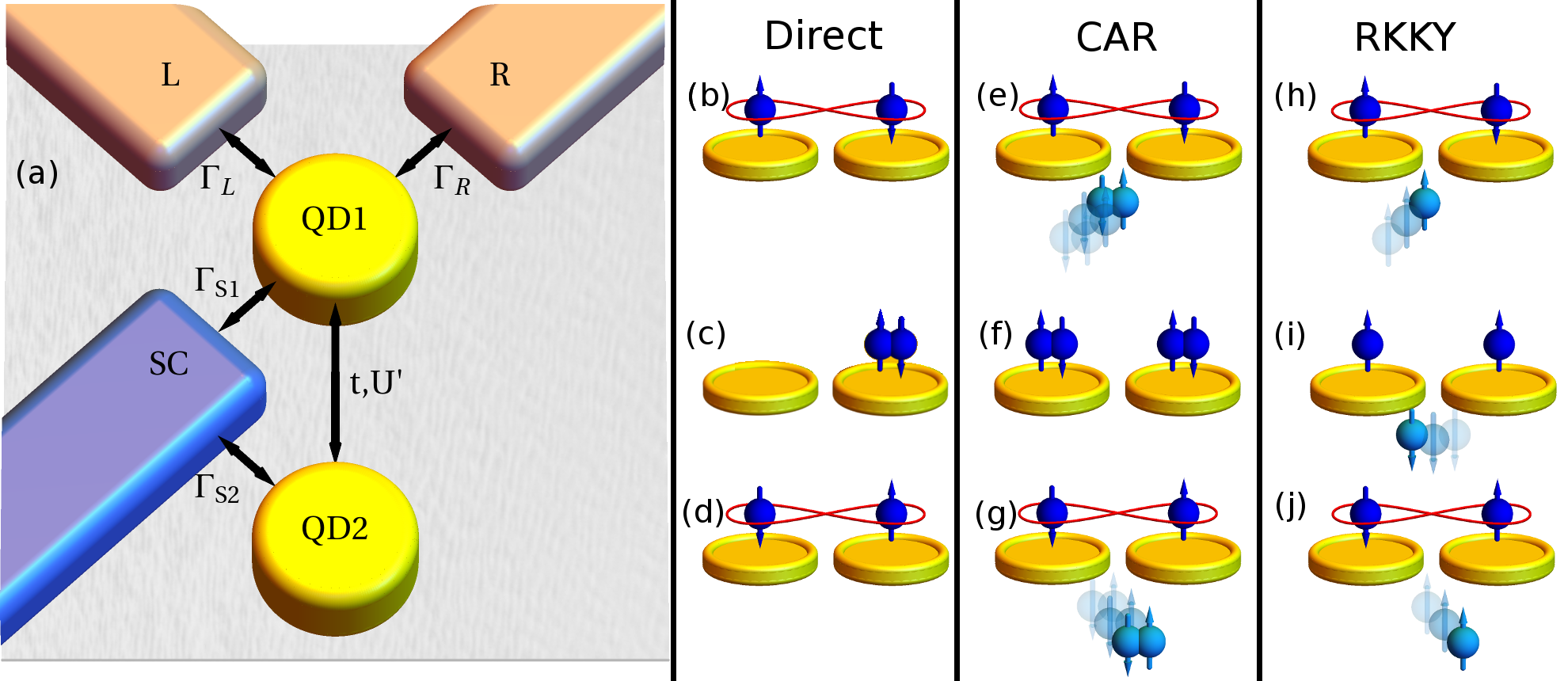}
\caption{
		 (a) Schematic of the considered system. Left/right (L/R) lead
		 is coupled to the first quantum dot (QD1), while superconductor
		 is attached to both QD1 and QD2.
		 (b)-(d) illustrate an example of direct spin exchange:
		 spin-up electron from the initial state (b) hops to the other QD (c) and spin-down electron 
		 hops back (d). Note, that the final state is in fact the same singlet state, 
		 only with opposite sign.
		 (e)-(g) show an example of process contributing to crossed Andreev reflection (CAR) exchange.
		 A Cooper pair from SC approaches DQD (e) and two singlets of the same charge 
		 are formed (f), before the Cooper pair is re-emitted (g).
		 (h)-(j) present an example of RKKY process: an electron scattered off
		 one QD (h) mediates the spin exchange towards the other (i), before it is finally scattered
		 off there, too (j).
		 }
\label{fig:system}
\end{figure}

In this paper we discuss the CAR-induced Kondo screening in a setup comprising T-shaped DQD
with normal and superconducting contacts, see \fig{system}(a).
We note that despite quite generic character of CAR exchange,
and its presence in systems containing at least two localized electrons
coupled close to each other to the same SC bath,
to best of our knowledge CAR-induced screening
has hardly been identified in previous studies
\cite{CPS1,CPS2,CPS4,CPS5,CPS9,IW_Kacper,IW_Sau,Zitko_Josephson,Zitko_S2QD,Martinek2017}.
In the system proposed here [\fig{system}(a)], its presence is evident.
Moreover, CAR exchange magnitude can be directly related to the relevant energy scales, such as the Kondo 
temperature, which provides a fingerprint for quantitative experimental verification of our predictions. 

The paper is organized as follows. In \Sec{model} we describe the considered system 
and present the model we use to study it. In \Sec{scales} the relevant energy scales are estimated
to make the discussion of main results concerning CAR-induced Kondo effect in \Sec{main} more clear. 
Finally, the influence of effects neglected in \Sec{main} are presented in the following sections,
including CAR exchange interplay with RKKY interaction (\Sec{RKKY}), particle-hole asymmetry (\Sec{asym}),
couplings asymmetry (\Sec{x}) and reduced efficiency of CAR coupling (\Sec{coef}). In summary,
the effects discussed in \Sec{main} remain qualitatively valid in all these cases.
The paper is concluded in \Sec{conclusions}.

\section{Model}
\label{sec:model}

The schematic of the considered system is depicted in \fig{system}(a).
It contains two QDs attached to a common SC lead.
Only one of them (QD1) is directly attached to the left (L) and right (R) normal leads,
while the other dot (QD2) remains coupled only through QD1.
The SC is modeled by the BCS Hamiltonian, 
$H_{\mathrm{S}}=\sum_{\vk\s}\xi_{\vk}a_{\vk\s}^{\dag}a_{\vk\s}-\Delta\sum_{\vk}(a^\dag_{\vk\up}a_{-\vk\down}^{\dag}+a_{-\vk\down}a_{\vk\up})$,
with energy dispersion $\xi_{\vk}$, energy gap $2\Delta>0$ and $a_{\vk\s}$ annihilation operator 
of electron possessing spin $\s$ and momentum $\vk$. The coupling between
SC and QDs is described by the hopping Hamiltonian 
$H_{\mathrm{TS}}=\sum_{i\vk\s}v_{\mathrm{S}i}(\dk_{i\s}a^{}_{\vk\s}+h.c.)$,
with $\dk_{i\s}$ creating a spin-$\s$ electron at QD$i$. The matrix element 
$v_{\mathrm{S}i}$ and the normalized density of states of SC in normal state, $\rho_{\rm S}$, 
contribute to the coupling of QD$i$ to SC electrode as $\GS{i} = \pi \rho_{\rm S} |v_{{\rm S}i}|^2$. 
We focus on the sub-gap regime, therefore, we integrate out SC degrees of freedom lying outside the energy gap \cite{RozhkovArovas}.
This gives rise to the following effective Hamiltonian,
$H_{\mathrm{eff}}=H_{\mathrm{SDQD}}+H_{\rm L}+H_{\rm R}+H_{\rm T}$, 
where 
\begin{eqnarray}
H_{\rm SDQD} 	\es 
				\sum_{i\s} \e_{i} n_{i\s} 
				+\sum_{i} U n_{i\up} n_{i\down} 
				+U' (n_1-1)(n_2-1) 
				\nn\\
				&+&\sum_\s t(\dk_{1\s}d^{}_{2\s} + h.c.) 
				+J \vec{S}_1\vec{S}_2
				\nn\\
				&+&\sum_{i} \!\!\left[ \Gamma_{{\rm S}i} (\dk_{i\up} \dk_{i\down} \!+\! h.c.)
				+\Gamma_{\rm SX} (\dk_{i\up} \dk_{\bar{i}\down} \!+\! h.c.) \right]
	\label{H_DQD} 
\end{eqnarray}
is the Hamiltonian of the SC-proximized DQD
\cite{IW_Kacper,Walldorf2018Feb}, with QD$i$ energy level $\e_i$,
inter-site (intra-site) Coulomb interactions $U'$ ($U$),
inter-dot hopping $t$, and CAR coupling $\GS{\rm X}$.
$n_{i\s}=\dk_{i\s}d^{}_{i\s}$ denotes the electron number operator 
at QD$i$, $n_i=n_\up+n_\down$, and $\bar{i}\equiv 3-i$. 
Our model is strictly valid in the regime where $\Delta$ is the largest 
energy scale. Nevertheless, all discussed phenomena are
present in a full model for energies smaller than SC gap.
Moreover, by eliminating other consequences of the presence of SC lead,
our model pinpoints the fact that the non-local pairing is 
sufficient for the occurrence of the CAR exchange.
The presence of out-gap states shall result mainly in additional broadening of DQD energy levels,
changing the relevant Kondo temperatures.
We note that the procedure of integrating out out-gap states neglects the 
RKKY interaction mediated by SC lead and other possible indirect exchange mechanisms%
 \footnote{
 Note, that by RKKY interaction we mean only such an effective exchange, 
 which arises due to multiple scattering of a single electron or hole, see \fig{system}(h)-(j).
 Other mechanisms leading to the total indirect exchange are considered separately.
 In particular, in the large gap limit, exchange described in Ref.~\cite{Yao} is in fact reduced to
 the CAR exchange, and additional antiferromagnetic contribution would arise for finite gap.
 }. 
To compensate for this,
we explicitly include the Heisenberg term $ J \vec{S}_1\vec{S}_2$ in
$H_{\rm SDQD}$, with $\vec{S}_i$ denoting the spin operator of QD$i$
and a Heisenberg coupling $J$ substituting the genuine RKKY exchange.

The normal leads are treated as reservoirs of noninteracting electrons,
$H_{r}=\sum_{\vk\s}\e_{r\vk}\ck_{r\vk\s}c^{}_{r\vk\s}$,
where $c^{}_{r\vk\s}$ annihilates an electron of spin 
$\s$ and momentum $\vk$ in lead $r$ ($r={\rm L,R}$) with the corresponding energy $\e_{r\vk\s}$.
The tunneling Hamiltonian reads,
$H_{\rm T} = \sum_{r\vk\s} v_{r} (\dk_{1\s}c^{}_{r\vk\s} + h.c.)$,
giving rise to coupling between lead $r$ and QD$i$ of strength $\Gamma_r = \pi \rho_r |v_r|^2$,
with $\rho_r$ the normalized density of states of lead $r$ and $v_r$ the 
local hopping matrix element, assumed momentum-independent.
We consider a wide-band limit, assuming constant $\Gamma_r=\Gamma/2$
within the cutoff $\pm D = \pm 2U$ around the Fermi level. 

For thorough analysis of the CAR exchange mechanism and its consequences
for transport, we determine the linear conductance between the two normal leads from
\begin{equation}
G = \frac{2e^2}{h} \pi \Gamma \int \left[ -\frac{\dd f_T}{\dd\w} \right] \A(\w) \de \w ,
\label{G}
\end{equation}
where $f_T$ is the Fermi function at temperature $T$,
while $\A(\w)$ denotes the normalized local spectral density 
of QD1 \cite{fn1}.
Henceforth, unless we state otherwise, we assume a maximal CAR coupling, 
$\GS{\rm X} = \sqrt{\GS{1}\GS{2}}$ \cite{IW_Kacper,Walldorf2018Feb},
$\GS{1}=\GS{2}=\GS{}$ and consider DQD tuned to the particle-hole symmetry point, 
$\e_1=\e_2=-U/2$. However, these assumptions are not crucial for the results presented
here, as discussed in Secs.~\ref{sec:asym}-\ref{sec:coef}.

\section{Estimation of relevant energy scales}
\label{sec:scales}

\begin{figure}
\includegraphics[width=1\linewidth]{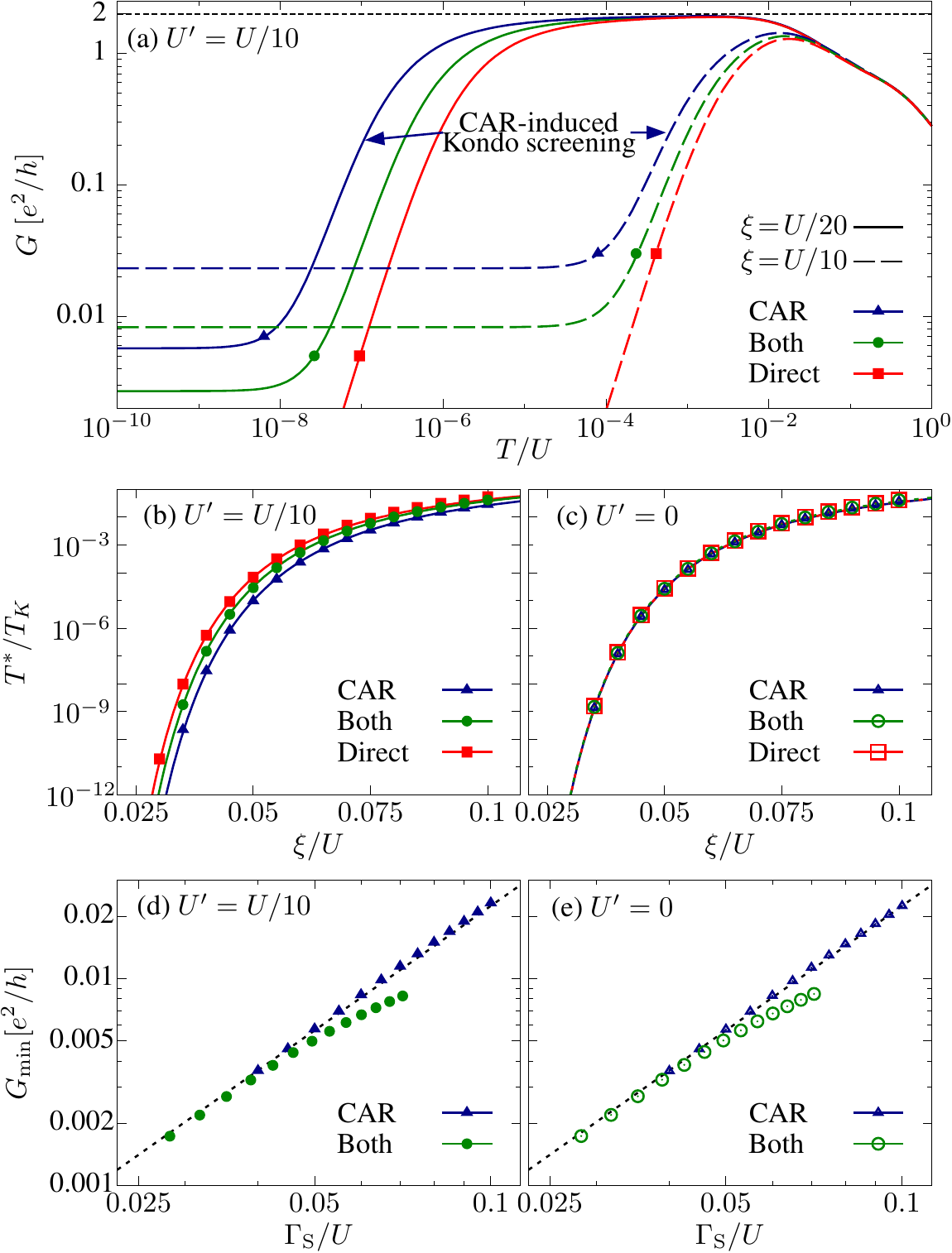}
\caption{(a) Linear conductance $G$ as function of $T$ calculated for 
		 $\e_1=\e_2=-U/2$, $\Gamma=U/5$, $U'=U/10$ and different situations, 
		 as indicated. The quantity $\xi\equiv\sqrt{\GS{}^2+t^2}$ is fixed 
		 for different curves drawn with the same dashing style.
		 Note the logarithmic scale on both axes.
		 (b) Points show $T^*/T_K$ calculated by NRG from curves in subfigure (a). 
		 Lines present the fit to \eq{Tstar} with $\Jeff$ obtained from \eq{Jeff}.
		 (c) The same as (b), only for $U'=0$.
		 (d) and (e) show the residual conductance $\Gmin \equiv G(T \!=\! 0)$ as a function of
		 $\GS{}$ for $t=0$ (denoted "CAR") and $t=\GS{}$ (denoted "Both"). 
		 Dotted line is a guide for eyes. $U'=U/10$ in (b) and (d) and $U'=0$ in (c) and (e).
		}
\label{fig:G-T}
\end{figure}

Since we analyze a relatively complex system, let us build up the understanding of its behavior starting
from the case of a QD between two normal-metallic leads, which can be obtained in our 
model by setting $t=\GS{}=J=U'=0$. Then, the conductance as a function of temperature, $G(T)$, grows
below the Kondo temperature $T_K$ and reaches maximum for $T\to 0$, $G(T\!=\!0)=G_{\rm max}$.
At particle-hole symmetry point, the unitary transmission is achieved, $G_{\rm max}= G_0 = 2e^2/h$;
see short-dashed line in \fig{G-T}(a).
An experimentally relevant definition of $T_K$ is that at $T=T_K$ 
$G(T)=G_{\rm max}/2$. $T_K$ is exponentially small in 
the local exchange $J_0 = 8\Gamma / (\pi \rho U)$, and is approximated by
$T_K \approx D \exp[-1/(\rho J_0)]$ \cite{Hewson_book}.

The presence of a second side-coupled QD, $t,U'>0$, significantly enriches the physics of the system 
by introducing direct exchange between QDs, see \fig{system}(b-d).
In general, effective inter-dot exchange can be defined as energy difference between 
the triplet and singlet states of isolated DQD, 
$\Jeff = E_{S=1} - E_{\rm GS}$. Unless $U$ becomes very large, superexchange can be neglected
\cite{Zitko_2QDEx} and $\Jeff$ is determined by \emph{direct exchange}, $\Jeff\approx 4t^2/(U-U')>0$.
When the hopping $t$ is tuned small \cite{CPS1}, one can expect $\Jeff\lesssim T_K$, which 
implies the two-stage Kondo screening \cite{Pustilnik_Glazman,Cornaglia}.
Then, for $T \ll T_K$, the local spectral density of QD1 serves as a band of width $\sim T_K$ for QD2.
The spin of an electron occupying QD2 
experiences the Kondo screening below the associated Kondo temperature
\begin{equation}
T^* = a T_K \exp(- b T_K / J_{\rm eff})
\label{Tstar}
\end{equation}
with $a$ and $b$ constants of order of unity \cite{Pustilnik_Glazman,Cornaglia}.
This is reflected in conductance, which drops to $0$ with lowering $T$, maintaining characteristic 
Fermi-liquid 
$G\sim T^2$ dependence \cite{Cornaglia}; see the curves indicated with squares 
in \fig{G-T}(a). Similarly to $T_K$, experimentally relevant definition of $T^*$ is that 
$G(T\!=\!T^*) = G_{\rm max}/2$. Even at the particle-hole 
symmetry point $G_{\rm max} < G_0$, because the single-QD strong-coupling fixed point 
is unstable in the presence of QD2 and $G(T)$ does not achieve $G_0$ exactly,
before it starts to decrease.

The proximity of SC gives rise to two further exchange mechanisms that
determine the system's behavior. First of all, the (conventional)
\emph{RKKY interaction} appears, $J \sim \GS{}^2$ \cite{RK,K,Y}. 
Moreover, the \emph{CAR exchange} emerges as a consequence of finite $\GS{}$ \cite{Yao}. 
It can be understood on the basis 
of perturbation theory as follows. DQD in the inter-dot singlet state may absorb
and re-emit a Cooper pair approaching from SC; see \fig{system}(e)-(g). As a second-order
process, it reduces the energy of the singlet, which is the ground state of isolated DQD.
A similar process is not possible in the triplet state due to spin conservation.
Therefore, the singlet-triplet energy splitting $\Jeff$ is increased (or generated for $t=J=0$). 
More precisely, the leading ($2$nd-order in $t$ and $\GS{}$) terms
in the total exchange are 
\begin{equation}
\Jeff 	\approx 	J + \frac{4t^2}{U-U'+\frac{3}{4}J} + \frac{4\GS{}^2}{U+U'+\frac{3}{4}J}.
\label{Jeff}
\end{equation}
Using this estimation, one can predict $T^*$ for finite $\GS{}$, $t$ and $J$ with \eq{Tstar}.
Apparently, from three contributions corresponding to:
(i) RKKY interaction, (ii) direct exchange and (iii) CAR exchange, only the first may bear a negative (ferromagnetic) sign.
The two other contributions always have an anti-ferromagnetic nature.
More accurate expression for $\Jeff$ is derived in Appendix~\ref{sec:downfolding}
[see \eq{A_J}] by the Hamiltonian down-folding procedure. The relevant terms differ 
by factors important only for large $\GS{}/U$. 
Finally, it seems worth stressing that normal leads are not necessary for CAR exchange to occur.
At least one of them is inevitable for the Kondo screening though, and two symmetrically coupled 
normal leads allow for measurement of the normal conductance.

It is also noteworthy that inter-dot Coulomb interactions
decrease the energy of intermediate states contributing to direct exchange 
[\fig{system}(c)], while increasing the energy of intermediate
states causing the CAR exchange [\fig{system}(f)].
This results in different dependence of corresponding terms in \eq{Jeff} on $U'$.
As can be seen in \figs{G-T}(b) and \ref{fig:G-T}(c), it has a significant effect 
on the actual values of $T^*$.

\section{CAR exchange and Kondo effect}
\label{sec:main}

To verify \eqs{Tstar}-(\ref{Jeff}) we calculate $G$ using
accurate full density matrix numerical renormalization group (NRG) technique \cite{WilsonNRG,Weichselbaum,FlexibleDMNRG,fn2}.
We compare $U'=0$ case with experimentally relevant value $U'=U/10$ \cite{Keller2013Dec}.
While for two close adatoms on SC surface RKKY interactions may lead to prominent consequences
\cite{Klinovaja}, the conventional ({\it i.e.} non-CAR) contribution should 
vanish rapidly when the inter-impurity distance $r$ exceeds a few lattice constants \cite{RKKYrange,SC_RKKY}. 
Meanwhile, the CAR exchange may remain significant for $r$ of the order
of coherence length of the SC contact \cite{Yao}. Therefore, we first neglect the conventional RKKY coupling and analyze its consequences in Sec.~\ref{sec:RKKY}.

The main results are presented in \fig{G-T}(a), showing the temperature dependence of $G$
for different circumstances. 
For reference, results for $\GS{}=0$ are shown, exhibiting 
the two-stage Kondo effect caused by \emph{direct} exchange mechanism.
As can be seen in \figs{G-T}(b) and \ref{fig:G-T}(c), an excellent agreement of $T^*$ found from NRG calculations and \eq{Tstar} 
is obtained with $a=0.42$ and $b=1.51$, the same for both $U'=0$ and $U'=U/10$. Note, 
however, that $\Jeff$ is different in these cases, cf. \eq{Jeff},
and $U'$ leads to increase of $T^*$.

Furthermore, for $t=0$ and $\GS{}>0$ the two-stage Kondo effect caused solely by the \emph{CAR
exchange} is present; see \fig{G-T}(a).
Experimentally, this situation
corresponds to a distance between the two QDs smaller than the superconducting coherence length,
but large enough for the exponentially suppressed direct hopping to be negligible.
While intuitively one could expect pairing to compete with any kind of magnetic ordering,
the Kondo screening induced by CAR exchange is a beautiful example of a superconductivity
in fact leading to magnetic order, namely the formation of the Kondo singlet.
This CAR-exchange-mediated Kondo screening is our main finding.
For such screening, \eq{Tstar} is still fulfilled with very similar 
parameters, $a=0.37$ ($a=0.35$) and $b=1.51$ ($b=1.50$) for $U'=0$ ($U'=U/10$),
correspondingly; see \figs{G-T}(b-c).
Moreover, as follows from \eq{Jeff}, $U'$ reduces CAR exchange, and therefore diminishes $T^*$.
For the same values of $\Jeff$, the dependence of $G(T)$ for $t=0$ and $\GS{}>0$ is hardly different 
from the one for $\GS{}=0$ and $t>0$ for $T\geq T^*$ (results not shown).
However, $G(T)$ saturates at residual value $\Gmin$ as $T\to 0$ only for finite
$\GS{}$, which at particle-hole symmetry makes $\Gmin$
the hallmark of SC proximity and the corresponding CAR exchange processes.
From numerical results, one can estimate it as
\begin{equation}
\Gmin = \frac{e^2}{h} \cdot c \, \frac{\GS{}^2}{U^2} 
	\qquad {\scriptstyle (\GS{1}=\GS{2}=\GS{})} ,
\label{Gmin}
\end{equation}
with $c\approx 2.25$, barely depending on $U'$ and getting smaller for $t>0$. 
This is illustrated in \figs{G-T}(d-e), where the dotted line corresponds to \eq{Gmin} with $c=2.25$. 

Lastly, in \fig{G-T}(a) we also present the curves obtained for $t=\GS{}$ chosen such, 
that the quantity $\xi=\sqrt{t^2+\GS{}^2}$ remains the same 
in all the cases.
This is to illustrate what happens when \emph{both} (direct and CAR) exchange interactions are
present. \fig{G-T}(c) clearly shows that $T^*$ remains practically unaltered for $U'=0$.
The comparison with \fig{G-T}(b) proves that in this case it practically does not depend 
on $U'$. The enhancement of direct exchange is compensated by the decrease of the CAR one. 
On the contrary, $\Gmin$ decreases for larger $t$ below the estimation given by Eq.~(\ref{Gmin}), 
as can be seen in \figs{G-T}(d-e). 

While analyzing the results concerning $\Gmin(\GS{})$ plotted in \figs{G-T}(d-e) 
one needs to keep in mind that $\Gmin$ is obtained at deeply cryogenic conditions. To illustrate
this better, $G(\GS{})$ obtained for $t=0$ and $T=10^{-6}U$ is plotted with solid line 
in \fig{3}. Clearly, for weak $\GS{}$ the system exhibits rather conventional (single-stage)
Kondo effect with $G=\Gmax\approx 2e^2/h$, while QD2 is effectively decoupled ($\Gmax<2e^2/h$
in the proximity of SC lead \cite{KWIW}). Only for larger values of $\GS{}$
the CAR exchange is strong enough, such that $T^*>T$ and the dependence $G(\GS{})$ continuously 
approaches the $T=0$ limit estimated by \eq{Gmin} and presented in \figs{G-T}(d-e).

\section{CAR-RKKY competition}
\label{sec:RKKY}

\begin{figure}
\includegraphics[width=0.98\linewidth]{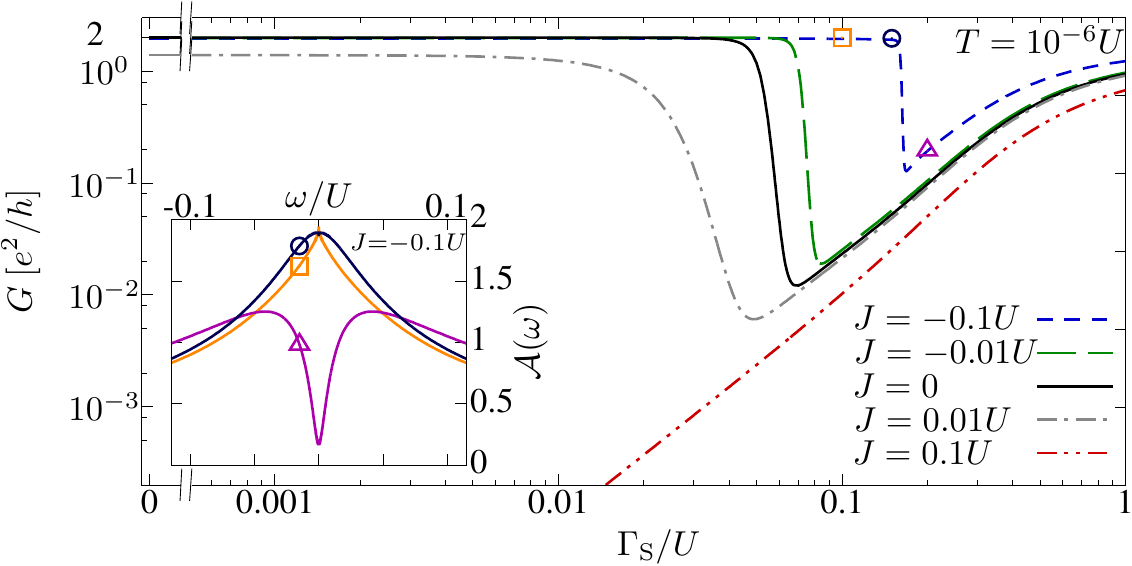}
\caption{Linear conductance $G$ vs. $\GS{}$ calculated
		 for $t=0$, $\Gamma=U/5$, $U'=U/10$, finite $T=10^{-6}U$
		 and different values of RKKY coupling $J$, as indicated. 
		 Inset shows QD1 spectral function $\mathcal{A}(\w)$ as a function of energy $\w$
		 for points on $J=-0.1U$ curve, indicated with corresponding symbols.
		}
\label{fig:3}
\end{figure}

Let us now discuss the effects introduced by the conventional RKKY interaction.
We choose $t=0$ for the sake of simplicity and
analyze a wide range of $\GS{}$, starting from the case of anti-ferromagnetic 
RKKY interaction ($J>0$). Large $J>0$ leads to the formation of a molecular singlet in the 
nanostructure. This suppresses the conductance, unless $\GS{}$ becomes of the order of $U/2$, 
when the excited states of DQD are all close to the ground state. This is illustrated 
by double-dotted line in \fig{3}.
Smaller value of $J>0$ causes less dramatic consequences, namely it just increases $\Jeff$ according
to \eq{Jeff}, leading to enhancement of $T^*$, cf. \eq{Tstar}. This is presented with
dot-dashed line in \fig{3}.

The situation changes qualitatively for ferromagnetic RKKY coupling, $J<0$.
Then, RKKY exchange and CAR exchange have opposite signs and compete with each other.
Depending on their magnitudes and temperature, one
of the following scenarios may happen.

For $\Jeff > 0$, {\it i.e.} large enough $\GS{}$, and $T<T^*$, the system is in the 
singlet state due to the two-stage Kondo screening of DQD spins. $G(T\!=\!0)$ is reduced 
to $\Gmin$, which tends to increase for large negative $J$; see dashed lines in \fig{3}. 
In the inset to \fig{3}, the spectral density of QD1 representative for this regime is plotted 
as curve indicated by triangle. It corresponds to a point on the $J=-0.1U$ curve in the main 
plot, also indicated by triangle. The dip in $\A(\w)$ has width of order of $T^*$.

For finite $T$, there is always a range of sufficiently small $|\Jeff|$, where QD2 becomes effectively
decoupled, and, provided $T<T_K$, $G$ reaches $\Gmax$ due to conventional Kondo effect 
at QD1. This is the case for sufficiently small $\GS{}$ for $J=0$ or $J=-0.01U$, and in the narrow
range of $\GS{}$ around the point indicated by a circle in \fig{3} for $J=-0.1U$ (for $J=0.01U$, 
the considered $T$ is close to $T^*$ and $G$ does not reach $G_{\rm max}$). The conventional Kondo effect manifests itself with 
a characteristic peak in $\A(\w)$, as illustrated in the inset in \fig{3} with line denoted by circle.

Finally, large enough $\Jeff < 0$ and low $T$, give rise to an effective ferromagnetic coupling of DQDs 
spins into triplet state. Consequently, the underscreened Kondo effect occurs 
\cite{Mattis,NozieresBlandin} for weak $\GS{}$ and, {\it e.g.}, $J=-0.1U$; 
see the point indicated by square in \fig{3}.
This leads to $G=\Gmax$ and a peak in $\A(\w)$, whose shape is significantly different from the
Kondo peak, cf. the curve denoted by square in the inset in \fig{3}.

\section{Effects of detuning from the particle-hole symmetry point}
\label{sec:asym}

\begin{figure}
\includegraphics[width=0.98\linewidth]{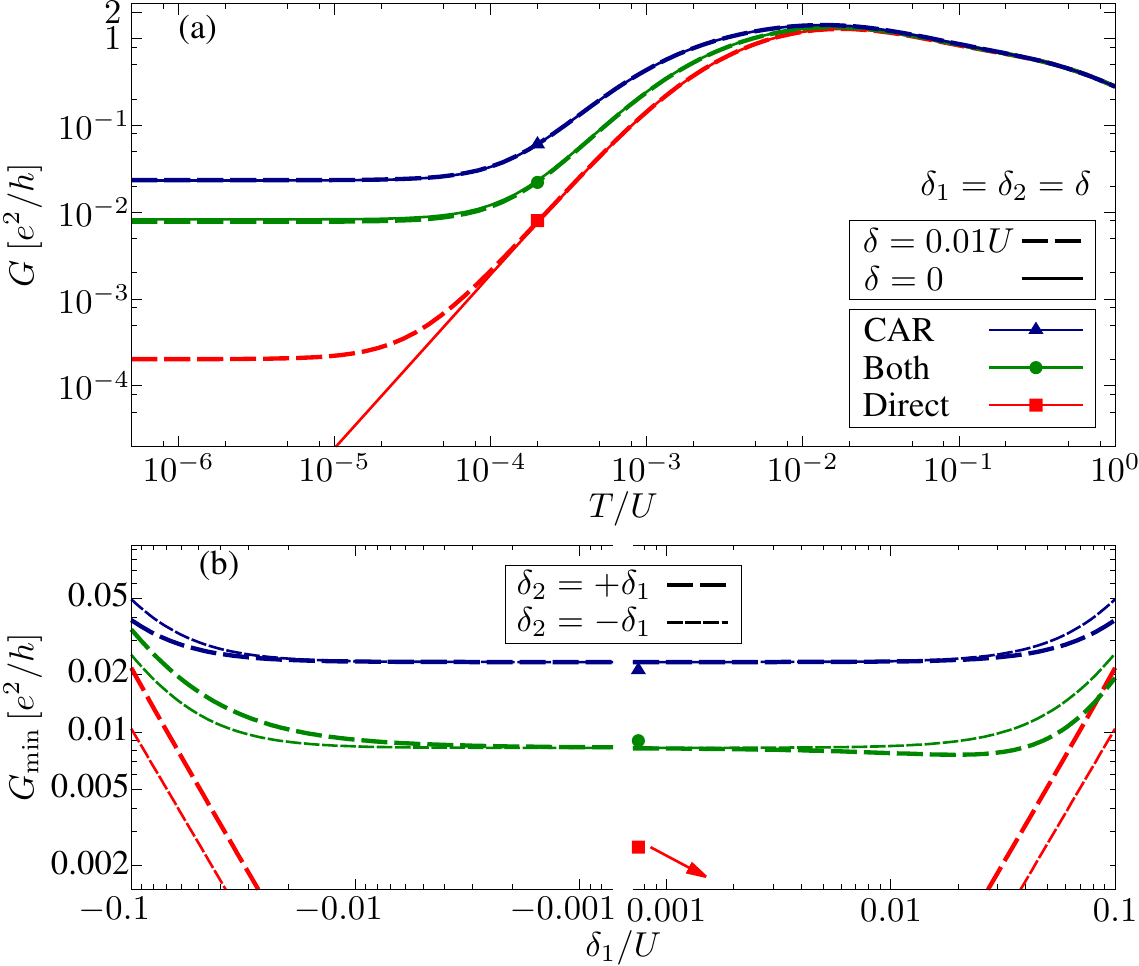}
\caption{
         (a) Linear conductance between the normal leads $G$ as a function of temperature $T$
         for parameters corresponding to \fig{G-T}(a) with $\xi=U/10$, and additional curves for finite 
         detuning from particle-hole symmetry point, $\delta_1=-\delta_2$, 
         and two values of $\xi=\sqrt{t^2+\GS{}^2}$, as indicated in the figure.
         (b) $\Gmin \equiv G(T \!=\! 0)$ as a function of QD1 detuning $\delta_1$ for different
         exchange mechanisms, $\xi=U/10$ and $\delta_2=\pm\delta_1$ (as indicated).
		}
\label{fig:asym}
\end{figure}

At PHS $\Gmin=G(T \!=\! 0)=0$ in the absence of superconducting lead, making $\Gmin > 0$ a hallmark
of SC-induced two-stage Kondo effect. However, outside of PHS point $\Gmin > 0$ even in the case of 
the two-stage Kondo effect caused by the direct exchange. 
Exact PHS conditions are hardly possible in real systems, and the fine-tuning of the QD energy
levels to PHS point is limited to some finite accuracy.
Therefore, there may appear a question, if the results obtained at PHS are of any importance for the
realistic setups. As we show below --- they are,
in a reasonable range of detunings $\delta_i=\e_i +U/2$.

In \fig{asym}(a) we present the $G(T)$ dependence in and outside the PHS, corresponding to 
parameters of \fig{G-T}(a). 
Clearly, for considered small values of $\delta_1=\delta_2=\delta$, 
$\Gmin<10^{-3}e^2/h$ for direct exchange only, while $\Gmin$ in the presence of a superconductor is 
significantly increased and close to the PHS value. Furthermore, for $|\delta_1| \sim |\delta_2| 
\sim \delta$, the residual conductance caused by the lack of PHS, $\Gmin \approx e^2/h \cdot (\delta/U)^2$,
which is a rapidly decreasing function in the vicinity of PHS point, as illustrated in \fig{asym}(b)
with lines denoted by a square. Evidently, in the regime $|\delta_i| < 0.01U$ the residual conductance
caused by SC is orders of magnitude larger, leading to the plateau in $\Gmin(\delta_1)$ dependence,
visible in \fig{asym}(b).
Taking into account that the realistic values of $U$ in the semiconductor quantum dots are rather 
large, this condition seems to be realizable by fine-tuning of QD gate voltages.

Lastly, let us point out that while in the presence of only one exchange mechanism, \emph{CAR} or
\emph{direct}, $\Gmin(\delta_1)$ dependencies depicted in \fig{asym}(b) are symmetrical with respect
to sign change of $\delta_1$, for \emph{both} exchange mechanisms the dependence is non-symmetric. 

\section{Effects of asymmetry of couplings to superconductor}
\label{sec:x}

\begin{figure}
\includegraphics[width=0.98\linewidth]{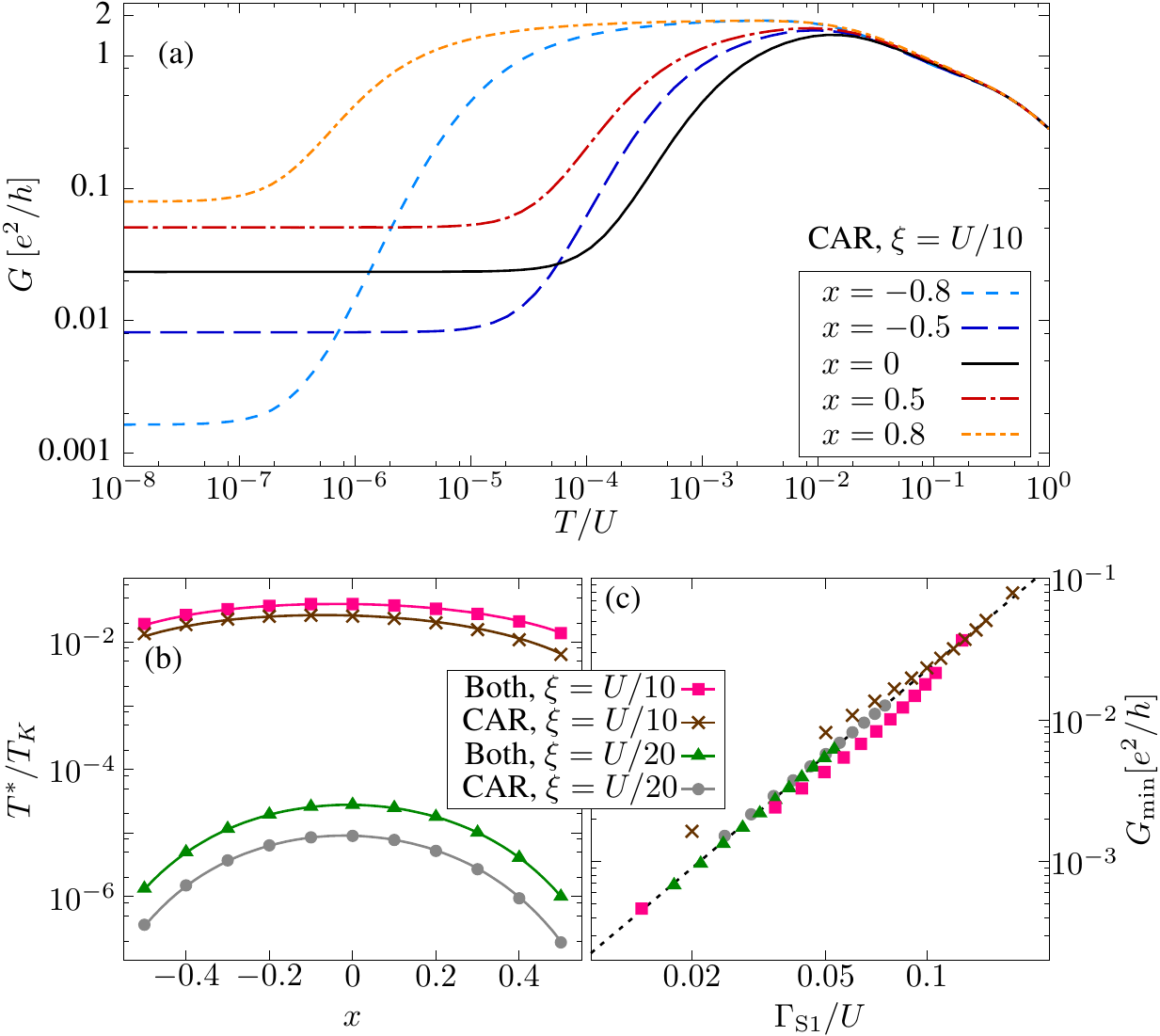}
\caption{
		 (a) Linear conductance between the normal leads, $G$, as a function of temperature, $T$,
		 for parameters corresponding to \fig{G-T}(a) with $\xi=U/10$, for different values 
		 of asymmetry coefficient $x$ [see \eq{xGS}], in the presence of \emph{CAR} exchange only.
		 (b) The second-stage Kondo temperature $T^*$ normalized by $T_K$ as a function of $x$, 
		 calculated with the aid of NRG (points) and a fit to \eq{Tstar} (lines) 
		 with $\Jeff$ from \eq{Jeff}.
		 (c) The zero-temperature conductance $\Gmin$ as a function of QD1 coupling to SC lead, $\GS{1}$,
		 compiled from data obtained at different circumstances (as indicated in the legend)
		 for different $x$. Dotted line corresponds to \eq{Gmin2} with $c=2.25$.
		}
\label{fig:x}
\end{figure}

Similarly to PHS, the ideal symmetry in the coupling between respective QDs and SC lead is hardly possible
in experimental reality. As shown below, it does not introduce any qualitatively new features.
On the other hand, it decreases the second stage Kondo temperature, which is already small, therefore,
quantitative estimation of this decrease may be important for potential experimental approaches.
To analyze the effects of $\GS{1}\neq\GS{2}$, we introduce the asymmetry parameter $x$ and extend
the definition of $\GS{}$,
\beq
x = \frac{\GS{1}-\GS{2}}{\GS{1}+\GS{2}}, \quad \GS{} = \frac{\GS{1}+\GS{2}}{2}.
\label{xGS}
\eeq
Note, that even for a fixed $\GS{}$, the actual CAR coupling $\GS{\rm X}=\GS{}\sqrt{1-x^2}$ decreases
with increasing $|x|$, which is a main mechanism leading to a decrease of $T^*$ outside the $x=0$ point
visible in \figs{x}(a) and (b). To illustrate this, the curves corresponding to \emph{both} exchange
mechanisms were calculated using $x$-dependent $t=\GS{\rm X}$ instead of $t=\xi/\sqrt{2}$. 
Therefore, $\xi$ was generalized for $x\neq 0$ by setting $\xi=\sqrt{t^2(1-x^2)^{-1}+\GS{}^2}$.
Clearly, in \fig{x}(b) the curves for different exchange mechanisms are very similar and differ mainly 
by a constant factor, resulting from different influence of $U'$; see \Sec{scales}. 
The magnitude of $T^*$ changes is quite large, exceeding an order of magnitude for $x=\pm 0.5$ 
and $\xi=U/20$. Moreover, $T^* \to 0$ for $x\to\pm 1$. Consequently, for strongly asymmetric
devices one cannot hope to observe the second stage of Kondo screening.

A careful observer can note that the $T^*(x)$ dependency is not symmetrical; note for example different 
$T^*$ for $x=\pm 0.5$ in \fig{x}(a). This is caused by the dependence of the first stage Kondo temperature
$T_K$ on $\GS{1}$ \cite{part1,DomanskiIW},
\beq
\widetilde{T}_K(\GS{1}) = T_K \cdot \exp\!\left( \frac{\pi}{2} \frac{\GS{1}^2}{\Gamma U}\right).
\eeq
Here, $T_K$ is, as earlier, defined in the absence of SC, while $\widetilde{T}_K$ is a function 
of $\GS{1}$, such that $G(\widetilde{T}_K) = G_{\rm max}(\GS{1})/2$ in the absence of QD2. 
As $\widetilde{T}_K$ grows for increasing $\GS{1}$ (or $x$), $T^*$ decreases according to \eq{Tstar}. 
Its $\GS{}$ dependence can be accounted for by small changes in the coefficients $a$ and $b$ in \eq{Tstar}, 
as long as $x$ is kept constant. 

To close the discussion of $T^*(x)$ dependence let us point out, that in \eq{A_J} 
there appears a correction to \eq{Jeff} for $x\neq 0$. However, it is very small due to additional
factor $\GS{}^2/U^2$ in the leading order. Its influence on curves plotted in \fig{x}(b) is hardly visible.

In turn, let us examine the $x$ dependence of the $T=0$ conductance $\Gmin$. As can be seen 
in \fig{x}(a), it monotonically increases with $x$, as it crosses $x=0$ point. In fact, \eq{Gmin}
can be generalized to
\beq
\Gmin = \frac{e^2}{h} \cdot c \, \frac{\GS{1}^2}{U^2} ,
\label{Gmin2}
\eeq
with $c\approx 2.25$ (indicated by a dotted line in \fig{x}(c)). Note that $\Gmin$ is proportional to 
$\GS{1}^2=(x+1)^2 \GS{}^2$, instead of simply $\GS{}$, cf. \eq{Gmin}. The values of $\Gmin$ obtained
from all analyzed $G(T)$ dependencies for different $x$ have been compiled in \fig{x}(c).
It is evident, that \eq{Gmin2} is approximately fulfilled for all the considered cases.

Finally, it seems noteworthy that the normal-lead coupling asymmetry, 
$\Gamma_{\rm L}\neq \Gamma_{\rm R}$, is irrelevant for the results except for a constant factor
diminishing the conductance $G$ \cite{KWIWJB-asym}.

\section{The role of CAR efficiency}
\label{sec:coef}

\begin{figure}[tb]
\includegraphics[width=0.98\linewidth]{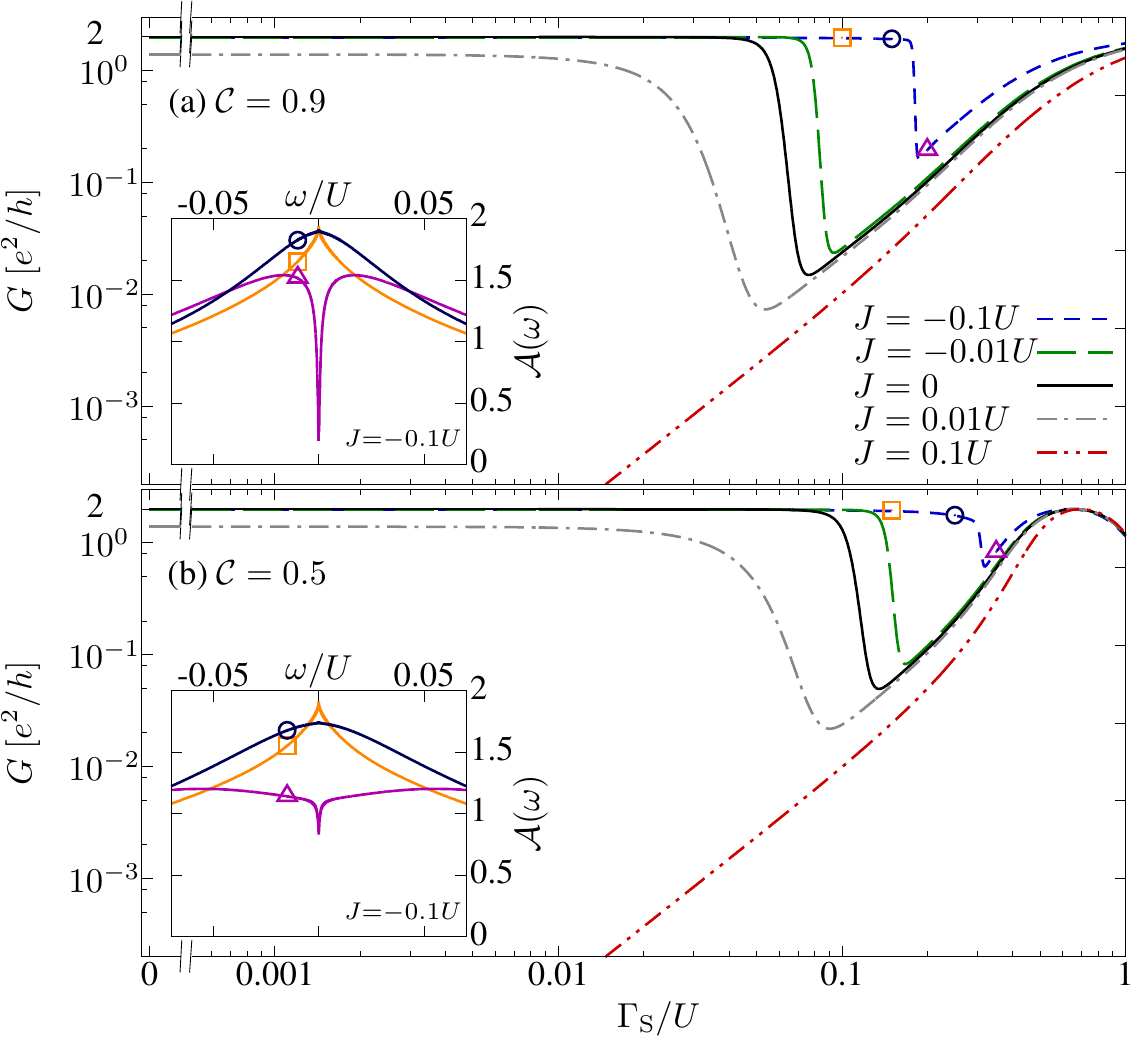}
\caption{Linear conductance between the normal leads
		 $G$ as a function of coupling to SC lead, $\GS{}$, for indicated values of RKKY exchange $J$
		 and the efficiency of CAR processes reduced by factor (a) $\coef=0.9$ and (b) $\coef=0.5$.
		 Other parameters as in \fig{3}.
		 Insets: QD1 local spectral density $\mathcal{A}(\w)$ as a function of energy $\w$
		 for points on $J=-0.1U$ curve, indicated with corresponding symbols.
		} 
\label{fig:C}
\end{figure}

Up to this point we assumed $\GS{\rm X} = \sqrt{\GS{1}\GS{2}}$, which is valid when the two 
quantum dots are much closer to each other than the coherence length in the superconductor.
This does not have to be the case in real setups, yet relaxing this assumption does not 
introduce qualitative changes. Nevertheless, the model cannot be extended to inter-dot 
distances much larger than the coherence length, where $\GS{\rm X}\to 0$.

To quantitatively analyze the consequences of less effective Andreev coupling we define the 
CAR efficiency as $\coef \equiv \GS{\rm X} / \sqrt{\GS{1}\GS{2}}$ and analyze $\coef < 1$
in the wide range of $\GS{1}=\GS{2}=\GS{}$ and other parameters corresponding to \fig{3}. 
The results are presented in \fig{C}.

Clearly, decreasing $\coef$ from $\coef=1$ causes diminishing of $\GS{\rm X}$, and consequently of CAR 
exchange. For a change as small as $\coef=0.9$, the consequences reduce to some shift of the 
conventional Kondo regime, compare \fig{C}(a) with \fig{3}. Stronger suppression of CAR may, 
however, increase the SC coupling necessary to observe the second stage of Kondo screening caused
by CAR outside the experimentally achievable range, see \fig{C}(b). Moreover, the reduced $T^*$
leads to narrowing of the related local spectral density dip, while the
increased critical $\GS{}$ necessary for the observation of the second stage of screening leads to the
shallowing of the dip. This is visible especially in the inset in \fig{C}(b).

\section{Conclusions}
\label{sec:conclusions}

The CAR exchange mechanism is present in any system comprising at least
two QDs or magnetic impurities coupled to the same superconducting contact
in a way allowing for crossed Andreev reflections.
In the considered setup, comprised of two quantum dots in a T-shaped geometry 
with respect to normal leads and proximized by superconductor,
it leads to the two-stage Kondo
screening even in the absence of other exchange mechanisms.
This CAR induced exchange screening is characterized by a residual 
low-temperature conductance at particle-hole symmetric case.
We have also shown that the competition between CAR exchange and RKKY
interaction may result in completely different Kondo screening scenarios.

The presented results bring further insight into the low-temperature
behavior of hybrid coupled quantum dot systems, which hopefully could be verified
with the present-day experimental techniques.
Moreover, non-local pairing is present also in bulk systems such as non-$s$-wave superconductors.
The question if an analogue of discussed CAR exchange may play a role there
seems intriguing in the context of tendencies of many strongly correlated materials
to possess superconducting and anti-ferromagnetic phases.

\begin{acknowledgments}
This work was supported by the National Science Centre in Poland through project no.
2015/19/N/ST3/01030.
We thank J. Barna\'{s} and T. Maier for valuable discussions.
\end{acknowledgments}


\appendix


\section{More precise estimation of the effective exchange}
\label{sec:downfolding}

We first briefly describe the Hamiltonian down-folding method, then we apply it to the model
considered in the main paper.

\subsection{General formulation of down-folding method}

Say we have the full Hilbert space of states divided into two sections, A and B. We think of 
A as being most important and of B as some addition, typically a set of high-energy states (in terms of the 
non-interacting part of the Hamiltonian). We can structure the secular equation for $H$ as
follows,
\beq
\left[\begin{array}{cc} 
	H_{\rm A} & H_{\rm AB}^\dagger \\
	H_{\rm AB} & H_{\rm B}
	\end{array}\right]
\left[\begin{array}{c} 
	\psi_{\rm A}  \\
	\psi_{\rm B}
	\end{array}\right] 
= E 
\left[\begin{array}{c} 
	\psi_{\rm A}  \\
	\psi_{\rm B}
	\end{array}\right].
\label{eq1}
\eeq
Treating it as a set of linear equations one can calculate components $\psi_{\rm B}$ as functions
of matrix elements of $H$, elements of $\psi_{\rm A}$ and the unknown eigenvalue $E$, 
$\psi_{\rm B} = (E-H_{\rm B})^{-1} H_{\rm AB} \psi_{\rm A}$. Thus, we obtain the equation
\beq
\left[H_{\rm A} + H_{\rm AB}^\dagger (E-H_{\rm B})^{-1} H_{\rm AB}\right] \psi_{\rm A} 
= E \psi_{\rm A},
\eeq
where the term in the square bracket can be called the effective Hamiltonian of the states A, 
$H_{\rm A}^{\rm eff}(E)$.
Note, that as long as $E$ on the left-hand-side is not approximated, this expression is exact,
nonlinear equation for eigenvalues $E$ of the original, full Hamiltonian and the projections 
of the full eigenstates $(\psi_{\rm A} \, \psi_{\rm B})^{\rm T}$ onto the space A.
In particular, if part A has a finite basis, one obtains the characteristic equation for $H$
eigenvalues by subsequently eliminating one state after another. In practice, 
however, one is usually interested in eliminating states of high energies and determination
of low-energy spectrum of the Hamiltonian, $E \ll H_{\rm B}$. Then, one can put $E\approx 0$.
In more general case, $E \approx \langle H_{\rm A} \rangle$ can be used, where 
$\langle H_{\rm A} \rangle$ is some kind of estimation of the relevant energy. 
The procedure is valid, as long as the energy dependence of $H_{\rm A}^{\rm eff}(E)$ does not 
influence its spectrum strongly, in particular for small interaction $H_{\rm AB}$.

\subsection{Application to the model under consideration}
\label{sec:downfolding2}

The spin singlet  $S=0$ subspace of the effective Hamiltonian for the superconductor-proximized double quantum dot
$H_{\rm SDQD}$ can be written in the form
\begin{widetext}
\beq
H_{\rm SDQD}^{S=0} = \left[
	\begin{array}{ccccc}
	-U-\frac{3}{4}J+\delta_1 + \delta_2 & t\sqrt{2} & t\sqrt{2} & \GS{\rm X}\sqrt{2} & -\GS{\rm X}\sqrt{2} \\
	t\sqrt{2} & 2\delta_1 -U' & 0 & \GS{1} & \GS{2} \\
	t\sqrt{2} & 0 & 2\delta_2 -U' & \GS{2} & \GS{1} \\
	\GS{\rm X}\sqrt{2} &  \GS{1} & \GS{2} & U' & 0 \\
	-\GS{\rm X}\sqrt{2} & \GS{2} & \GS{1} & 0 & 2(\delta_1+\delta_2)+U' 
	\end{array}
	\right],
\label{H_S0}
\eeq
\end{widetext}
with $\delta_i = \e_i +U/2$; see \eq{H_DQD}. The order of basis states is as follows: 
$\ket{1}=2^{-1/2}(\ket{\up\down}-\ket{\down\up})$,
$\ket{2}=\ket{20}$, $\ket{3}=\ket{02}$, $\ket{4}=\ket{00}$, $\ket{5}=\ket{22}$,
where $\ket{\chi_1 \chi_2}$ denotes such state, that QD$i$ is in a state $\ket{\chi_i}$, and 
the possible states are $\chi_i=0$ (empty QD), $\chi_i=2$ (doubly occupied QD), or $\chi_i=\s$
(QD occupied by a single electron of spin $\s$; $\s=\up$ or $\s=\down$).

From \eq{H_S0} one can clearly see that in the regime of $U$ dominating over other energy scales
there is one state possessing energy of the order of $-U$, and other states have energies regular
in the limit $U \to \infty$. Therefore, the first approximation to $E_{\rm GS}$ is the energy of that
state, $E_{\rm GS}^0 = -U -\frac{3}{4}J+\delta_1 + \delta_2$. 
The down-folding procedure shall be used to correct this estimation. This correction is 
crucial, since otherwise we get an obvious yet crude result $\Jeff = J$.

Let us first examine $\GS{1}=\GS{2}=\GS{\rm X}=\GS{}=0$ case. 
Then, the charge is conserved, so the states $\ket{4}$ 
and $\ket{5}$ are decoupled. Taking $\ket{1}$ to subspace A, while keeping $\ket{2}$ and $\ket{3}$
in subspace B one gets the estimation of $E_{\rm GS}$, which for $E$ substituted by $E_{\rm GS}^0$ 
leads to 
\beq
\Jeff_{\GS{}=0} = J + 4\frac{t^2}{U-U'} 
						\left[ 1- \frac{(\delta_1-\delta_2)^2}{(U-U')^2} \right]^{-1},
	\label{A_Jt}
\eeq
in agreement with Refs.~\cite{Cornaglia,Ferreira}.

For finite $\GS{1}$, $\GS{2}$ and $\GS{\rm X}$ the situation is more complicated,
because there are four states to be eliminated.
We assign them all to subspace B and obtain cumbersome expressions, resulting from inverting 
explicitly non-diagonal $4\times 4 $ matrix. Therefore, we limit ourselves to the case 
of particle-hole symmetry, $\delta_1=\delta_2=0$. 
Then, keeping only $\ket{1}$ in subspace A and the remaining $4$ states in B, 
and using $E \mapsto E_{\rm GS}^0$ substitution one gets feasible result,
\beqa
\Jeff
	\es
	J 
	+\frac{4\GS{\rm X}^2}{U+U'+\frac{3}{4}J}\left[1-\frac{4x^2 \GS{}^2}{(U+\frac{3}{4}J)^2 -U'^2}\right]^{-1}
	\nn\\&&
	+\frac{4t^2}{U-U'+\frac{3}{4}J}\left[1-\frac{4\GS{}^2}{(U+\frac{3}{4}J)^2 -U'^2} \right]^{-1}
	\!\!\!\!\! . \qquad 
	\label{A_J}
\eeqa
Expanding \eq{A_J} in powers of $\GS{\rm X}$, $\GS{}$ and $t$, we obtain in the leading order
\beq
\Jeff 
	\approx 
	J+\frac{4\GS{\rm X}^2}{U+U'+\frac{3}{4}J}+\frac{4t^2}{U-U'+\frac{3}{4}J},
\label{A_J-final}
\eeq
identical to \eq{Jeff}, yet extended to $\GS{1}\neq \GS{2}$ and $\GS{\rm X} < \sqrt{\GS{1}\GS{2}}$;
see \Sec{coef} for discussion of the importance of the latter.


\begin{thebibliography}{67}
\bibitem [{fn1()}]{fn1}%
  \BibitemOpen
  \href@noop {} {\bibinfo  {journal} {$\mathcal{A}(\w)$ is given by
  $\mathcal{A}(\w) = -\Gamma \;\Im \GF{d^{}_{1\s} ; \dk_{1\s}}^r(\w)$, where
  $\GF{d^{}_{1\s} ; \dk_{1\s}}^r(\w)$ is the Fourier-transformed retarded
  Green's function of QD1}\ }\BibitemShut {NoStop}%
\end{thebibliography}
%

\end{document}